\documentclass[prl, aps,twocolumn,showpacs]{revtex4}

\usepackage{times}
\usepackage{amsmath}
\usepackage{amssymb}
\usepackage{amsthm}
\usepackage{graphicx}  
\usepackage{dcolumn}   
\usepackage{bm}        
\usepackage{multirow}
\usepackage{mathrsfs}
\usepackage{psfrag}


\newcommand{\sign}{\text{sign}}

\begin{document}

\author{Earl T.\ Campbell} \email{earl.campbell@materials.ox.ac.uk}
\author{Joseph Fitzsimons }
\author{Simon C.\ Benjamin}
\author{Pieter Kok}
\affiliation{Quantum \& Nano Technology Group, Department of  Materials, Oxford University, Oxford, UK}

\date{\today}
\title{Adaptive strategies for graph state growth in the presence of  monitored errors}

\begin{abstract}
\noindent
Graph states (or cluster states) are the entanglement resource that  enables one-way quantum
computing.  They can be grown by projective measurements on the
component qubits. Such measurements typically carry a significant
failure probability. Moreover, they may generate
imperfect entanglement. Here we describe strategies to adapt growth
operations in order to cancel incurred errors. Nascent states that
initially deviate from the ideal graph states evolve toward the
desired high fidelity resource without impractical overheads.
Our analysis extends the diagrammatic language of graph
states to include characteristics such as tilted vertices, weighted
edges, and partial fusion, which arise from experimental
imperfections. The strategies we present are relevant to
parity projection schemes such as optical `path erasure' with
distributed matter qubits.

\end{abstract}
\maketitle

\noindent
Graph states have the remarkable property that they embody all the  entanglement needed for quantum algorithms. The computation then  proceeds purely through single-qubit measurements, consuming
the graph state as a resource \cite{RB01a, RBB01a, HEB01a}. Several  physical mechanisms
that can create graph states have been identified, many of which  employ measurements in order to create the required entanglement \cite {BK01a,LBK01a,LBBKK01a,BR02a,BES01a,BBFM01a,Browne}. Efficient graph  state creation is possible even
when these entangling measurements have a high failure probability,
provided that success is heralded \cite {KLM01a,YoranReznik,N01a,BR02a, KMMRM01a, BBFM01a, DR01a, GKE01a}.  Failure corresponds, at
worst, to local (repairable) damage to the growing graph state.
This approach has been applied to linear optical scenarios, and to
scenarios involving macroscopically separated matter qubits.
A successful entangling measurement must have a {\em high
  fidelity}: Imperfect (non-maximal) entanglement generally leads to
errors in the computation. However, achieving a higher fidelity by
enforcing more stringent success criteria will generally result in a  large
resource overhead \cite{DR01a}.

We show that by adaptively altering the growth process
of graph states, there is a class of imperfections that can be
tolerated in creating ideal graph states.  We consider {\em monitored
errors}, i.e. random errors that cannot be predicted but which are
known once they have occurred. Such errors may originate from  frequency mismatch, spatial mode mismatch, or cavity coupling  mismatch. Here, we illustrate our techniques by considering the non- maximal entanglement that occurs when two sources in a path erasure  scheme have unequal photon emission rates \cite{Bose,Feng,Cabrillo}:  A spontaneously emitted photon that is detected early in the  detection window is more likely to have originated from the more  rapidly emitting cavity; conversely, photon detection late in the  window implies a bias to the slower cavity.  Rather than abandoning  such events as failures, we exploit the fact that the resulting  entanglement is a known function of the detection variable (e.g., the  observation time). Non-ideal measurements are employed to create  ideal graph states, dramatically reducing the resource overhead.

\begin{figure}[t]
    \centering   
    \includegraphics{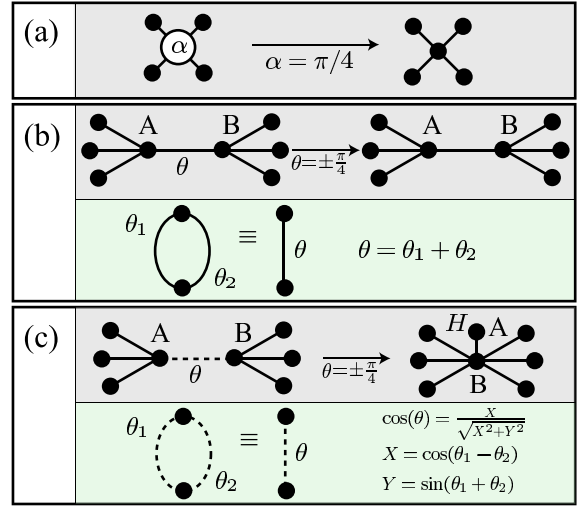}
    \caption{Graph generalizations (left), the corresponding proper  graphs (right), and the relevant addition rules. (\textbf{\textbf {a}}) tilted vertices are hollow circles labelled $\alpha$, denoting  a qubit in a state $\cos( \alpha) \vert 0 \rangle + \sin (\alpha)  \vert 1 \rangle$ prior to applying two qubit operations; (\textbf {\textbf{b}}) weighted graph edges, illustrated as a solid edge  between two qubits $A$ and $B$, and labelled $\theta$.  In operator  notation a weighted edge is $U_{AB}(\theta)= \cos (\theta) \openone +  i \sin (\theta) Z_{A}Z_{B}$; \textit{(\textbf{\textbf{c}})} partial  fusions, illustrated as a dashed line between two qubits $A$ and $B$  labelled $\theta$.  In operator notation a partial fusion is $P_{AB} (\theta)= \cos (\theta) \openone + \sin (\theta)Z_{A}Z_{B}$.}
    \label{Identities}
\end{figure}

In order to incorporate the effects of monitored errors, we extend  the graph state formalism by introducing {\em tilted vertex  amplitudes}, {\em weighted graph edges} \cite{DHLB01a, APDVB01a}, and  {\em partial fusions} (see
Fig.~\ref{Identities}). Tilted vertices arise directly from the  monitored errors, whereas weighted edges and partial fusions may  result from measurements on tilted vertices.  This broader class of  multi-qubit states still has a graphical description whose complexity  increases only polynomially with the number of qubits. We present  three adaptive growth strategies that yield ideal graph states in the  presence of monitored errors; we will refer to these as realignment,  merging and bridging.

\begin{figure*}[t]
    \centering   
    \includegraphics{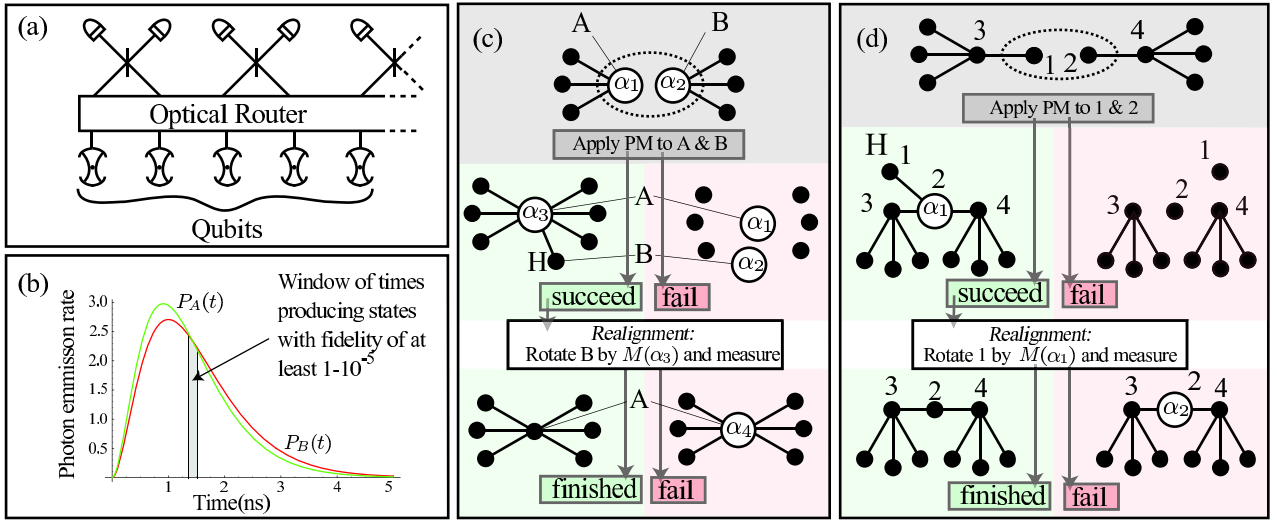}
    \caption{({\textbf{a}}) A distributed quantum computer; the qubits  are stored in matter systems which are coupled to optical modes.  Pairs of optical paths are routed towards a beam splitter while two  detectors (a PM device).  ({\textbf{b}}) emission rates from two  mismatched cavities; detection events within the shaded region  correspond to high-fidelity entanglement. ({\textbf{c}}) fusing  tilted GHZ states by a projective measurement. Upon success, this  will result in a larger tilted GHZ state.  This can be  probabilistically purified into a proper GHZ state. (\textit{\textbf {d}}) fusing two cherries generates a tilted central vertex with a  single cherry (which is used for realignment).  If realignment fails  the state can be used for merging  (Fig.~(\ref{merge}a)) or bridging  (Fig.~(\ref{bridge}a)). }
    \label{GHZfuse}
\end{figure*}

We will consider schemes for graph state construction that use
Projective Measurements (PMs) to construct graph states.  A PM is a
probabilistic entangling operation that, when successful to a high
fidelity, results in a projection of two qubits onto the odd parity
subspace.  An example of such a PM scheme is
given in Ref.~\cite{BK01a}. In this proposal matter qubits are  initially prepared in the state $ \vert + \rangle
\equiv (\vert 0 \rangle + \vert 1 \rangle)/\sqrt{2}$.  The qubits
emit photons depending on their state, and which-path erasure is used  to create two-qubit entanglement (Fig.~\ref{GHZfuse}a).  In the ideal
situation, when only a single detector clicks, the qubits are
projected onto the maximally entangled $(\vert 01 \rangle + \vert 10
\rangle)/\sqrt{2}$ state.  However, with partially distinguishable
photon sources, the qubits are projected onto a state of the form, $
\cos (\alpha) \vert 01 \rangle + \sin (\alpha) \vert 10 \rangle$
\cite{CFBK01a}.  If the distinguishibility results from the two qubits
having different photon emission rates, see Fig.~(\ref{GHZfuse}b),
then knowing the time of the detector click means the value of
$\alpha$ is also known, and hence monitored.  With current technology  detectors can resolve time several orders of magnitude faster than  typical cavity emission times, allowing accurate error monitoring.  Thus a dominant error source may be subsumed into our graphical  language by the introduction of tilted vertices.

\paragraph{Tilted vertices} A tilted vertex is parametrised by an  angle $\alpha$, which defines the initialisation state of that qubit  as $\vert \psi \rangle = \cos(\alpha) \vert 0 \rangle + \sin (\alpha)  \vert 1 \rangle$.  When the parameter is $\alpha=\pi/4$ the qubit is  a proper vertex, as illustrated in Fig.~(\ref{Identities}a).   Additional graph generalizations will be shown to arise when certain  measurements are made on tilted vertices.

\paragraph{Weighted edges} A weighted graph edge between two vertices  $A$ and $B$ is defined in operator notation as $U_{AB}(\theta) = \cos  (\theta) \openone + i \sin (\theta) Z_{A}Z_{B}$, and is graphically  represented as a solid edge labelled with an angle $\theta$.  The  angle is constrained to the range $- \pi/4 \leq \theta \leq \pi / 4$,  by using the identity $U_{AB}(\theta+ \pi/2)= i Z_{A}Z_{B}U_{AB} (\theta)$.  A weighted edge $U_{AB}(\theta)$ is local unitary  equivalent to a control-$Z(4 \theta)$; where $Z(\varphi)$ is the  diagonal matrix with elements $(1, e^{i \varphi})$.  For brevity  these local equivalences will be omitted, such that weighted edges  with $\theta = \pm \pi /4 $ are equivalent to a control-$Z$,  represented by a proper graph edge, as illustrated in Fig.~(\ref {Identities}b).

\paragraph{Partial fusions} A partial fusion between two qubits $A$  and $B$ is defined in operator notation as $P_{AB}(\theta) = \cos  (\theta) \openone + \sin (\theta) Z_{A}Z_{B}$, and is graphically  represented by a dashed line labelled $\theta$.  The angle is again  constrained to a $\pi/2$ range, by the identity $P_{AB}(\theta + \pi/ 2) =Z_{A}Z_{B} P_{AB}(-\theta)$.  When $\theta = + \pi/4$ or $\theta  = - \pi/4$, the operator becomes a projector onto the even or odd  parity subspace, respectively, as occurs with type-II fusion \cite {BR02a}.  For a full fusion $P_{AB}(\pm \pi /4)$ on proper vertices, the  resulting state is equivalent to a pure graph state, as in Fig.~(\ref {Identities}c).  This last required graph generalization differs from  the previous two in its non-unitary nature, and hence there will be  an implicit renormalization in all expressions.

We now describe strategies for adapting graph state synthesis in  response to monitored errors. The  microcluster approach~\cite{N01a}  is suitable as the overall scheme: GHZ states are created and  selectively fused together in order to produce any desired tologopy.  We begin by noting that a simple {\em realignment} process allows us  to correct a tilted vertex through the sacrifice of a neighbour.  We  then observe that, even without such neighbours, a proper graph can  still be constructed using strategies we call {\em merge} and {\em  bridge}. The steps involved are probabilistic, but upon failure the  latter strategies leave residual entanglement which can be exploited  in subsequent attempts. The merge procedure is efficient at  increasing the size of medium sized GHZ states, while the bridge  procedure can create edges between multi-neighbour nodes. Thus these  two strategies would be relevant at different stages of the  microcluster growth scheme, for example.

\paragraph{Realignment} The realignment strategy is most clearly  described using GHZ states.  A proper GHZ state is any proper graph  that has no more than one node with multiple neighbours, or any local  unitary equivalent graph.  The vertex with many neighbours is called  the \textit{core} vertex, and the attached single neighbour vertices  are its \textit{cherries}.  Generalized GHZ states can be constructed  by projectively measuring the core qubits of two smaller generalized  GHZ states, as illustrated in Fig.~(\ref{GHZfuse}c). If successful,  the state has one tilted vertex to be corrected.  The value of $ \alpha_{3}$ will be determined by $\alpha_{1}$, $\alpha_{2}$, and the  photon detection time; increasing the expected amount of entanglement  associated with $\alpha_{3}$ is discussed in \cite{CFBK01a}.  A  realignment strategy, shown in Fig.~(\ref{GHZfuse}c), can correct the  tilted vertex by measuring a cherry in a basis tuned to $\alpha_{3}$.  Note that a tilted vertex could not possibly be corrected through  purely local operations on that vertex, since the tilt implies  incorrect entanglement relations with the neighbouring vertices. The cherry qubit is rotated by $M_{B}(\alpha_{3})$, where $M(\alpha)= \sin  (\alpha) X - \cos (\alpha) Z$, which becomes $H$, the Hadamard, for $ \alpha=-\pi/4$. With success probability $p_{s}(\alpha_{3})=\frac{1} {2}\sin^{2}(2\alpha_{3})$ the tilting will be removed.  If the  realignment is unsuccessful, then the tilting is exacerbated such  that its angle changes to $R(\alpha_{3}) =  \arccos (\cos^{2}(\alpha_3)(1- p_{s} (\alpha_3))^{\frac{-1}{2}})$.  Note that the realignment procedure is not  specific to tilted vertices in generalized GHZ states, but can be  used to correct any tilted vertex that has a cherry.  A less risky  procedure, which constructs graph states more complex
than GHZ states, requires a PM on two cherries (Fig.~\ref{GHZfuse}d).  The risk is less because a failure results in
only two qubits being separated.  When successful one of the qubits
becomes a tilted \textit{intercore} vertex, and the other becomes its
cherry.  Again the cherry can be used for realignment.  Through  repeated applications of this procedure one could realize any graph  topology, including the cubic lattice graphs known as cluster states. 

\begin{figure}[t]
    \centering
    \includegraphics{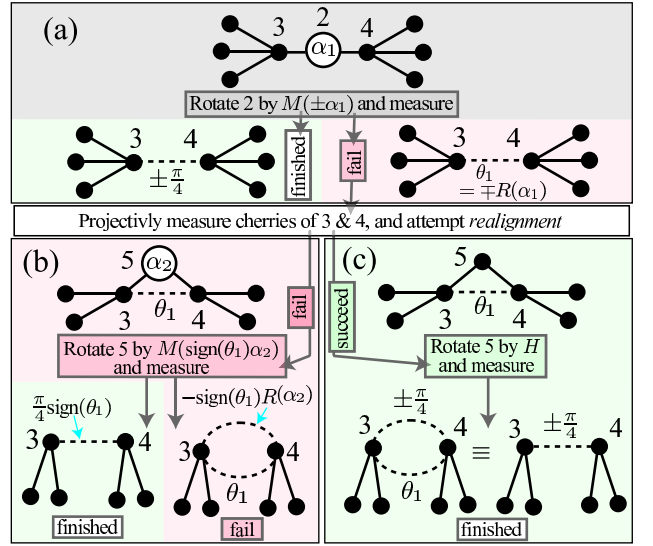}
    \caption{Connecting two subgraphs by merging one qubit from each  subgraph.  The qubits are labelled by $3$ and $4$, and a successful  merger corresponds to $P_{34}(\pm \pi/4)$. (\textit{\textbf{a}}) an  attempt to merge two qubits attached to a tilted vertex with success  outcome $P_{34}(\pm \pi/4)$ or failure outcome $P_{34}(\mp R(\alpha_ {1}))$. After a failure, projective measurements must be repeated  until two more cherries are fused.  A realignment attempt can then be  made on the resulting tilted vertex by measuring its only { \em  cherry}.  If realignment is successful then implement (\textit{c}),  but if failed then implement (\textit{b}). (\textit{\textbf{b}}) is a  probabilistic procedure that on success generates $P_{34}(\pm \pi)$  and on failure $P_{34}(\mp R(\alpha_{2}))$.  (\textit{\textbf{c}}) is  a deterministic procedure that projects with $P_{34}(\pm \pi/4)$.}
    \label{merge}
\end{figure}

\begin{figure}[t]
    \centering
    \includegraphics{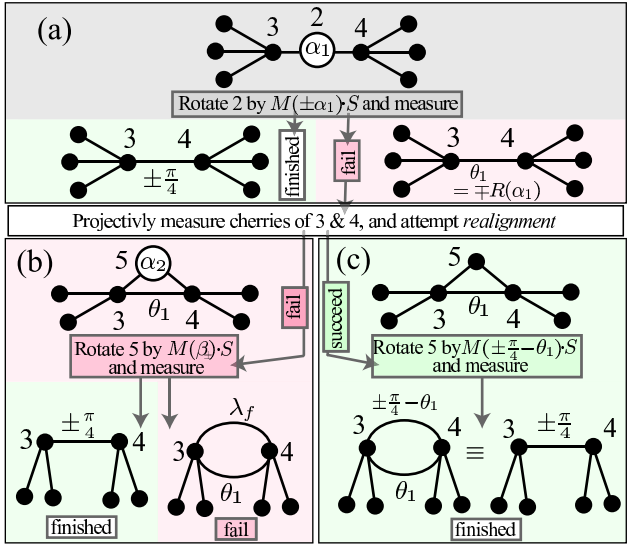}
    \caption{Connecting two subgraphs by bridging one qubit from each subgraph.  The qubits are labelled by $3$ and $4$, and a successful  bridge corresponds to $U_{34}(\pm \pi/4)$. (\textbf{\textit{a}}) an  attempt to bridge two qubits attached to a tilted vertex with success  outcome $U_{34}(\pm \pi/4)$ or failure outcome $U_{34}(\mp  R(\alpha_ {1}) )$. After a failure, projective measurements must be repeated  until two more cherries are fused.  A realignment attempt can then be  made on the resulting tilted vertex by measuring its only cherry.  If  realignment is successful then implement (\textit{c}), but if it  fails then implement (\textit{b}). (\textit{\textbf{b}}) is a  probabilistic procedure that on success generates $U_{34}(\pm \pi/4 - \theta_{1})$ and on failure $U_{34}(\lambda_{f})$, where $\beta_{\pm}$ and  $\lambda_{f}$ are respectively defined by (\ref{BETA}) and (\ref {LF}).  (\textit{\textbf{c}}) is a deterministic procedure that  projects with $U_{34}(\pm \pi/4 -\theta_{1})$.}
            \label{bridge}
\end{figure}

The remainder of this letter concerns what use can be made of this
tilted two-neighbour vertex after all of its cherries have been lost.
If the tilted vertex is connected to two vertices labelled $3$ and
$4$, then two options are available: (\textit{i}) to attempt to
\textit{merge} $3$ and $4$ with $P_{34}(\pm \pi/4)$, as in
Fig.~(\ref{merge}); (\textit{ii}) to attempt to \textit{bridge} $3$
and $4$ with $U_{34}(\pm \pi/4)$, as in Fig.~(\ref{bridge}).

\paragraph{Merging} The protocol for the first attempt at merger is  shown in Fig.~(\ref{merge}a).  Note that, if the vertex was not  tilted, then this can be deterministically achieved by measuring the  intercore vertex in the $X$-basis.  However, when the vertex is  tilted the procedure becomes probabilistic.  Either the even $P_{34} (\pi/4)$ or odd $P_{34}(-\pi/4)$ parity projector can be targeted by  rotating by $M_{2}(\alpha_{1})$ or $M_{2}(-\alpha_{1})$, respectively.   Success occurs with probability $p_{s}(\alpha_{1})=\frac{1}{2} \sin^ {2}(2 \alpha_{1})$.  Failure creates a partial fusion onto the  subspace that is orthogonal to the targeted subspace; hence the sign  flipping of $P_{34}(\mp R(\alpha_{1}))$.

The entanglement from this partial fusion will increase the  probability of success for subsequent attempts at merging.  For  another attempt to be made at merging, first a successful PM must be  achieved.  This will generate a new tilted intercore vertex with a  cherry that can be used in a realignment attempt.  If realignment is  successful, then Fig.~(\ref{merge}c) shows how this can be used to  deterministically add to the partial fusion to get a full parity  projection, which only differs from the untilted case by skewing the  probabilities of the even or odd projection results.  If realignment  is unsuccessful, then Fig.~(\ref{merge}b) shows how a probabilistic  attempt can be made at merging.  The probability of success is $p_{m}(\alpha_{2},\theta_{1})=p_{s}(\alpha_{2})(1 \pm \sin(2 \theta_{1}))$, where the sign freedom, $\pm$, comes from the choosen rotation $M(\pm \pi/4)$.  Hence, the partial entanglement will always be beneficial if we match $\pm$ to the sign of $\theta_{1}$.  If this attempt at merging fails then it  causes an additional partial fusion of $P_{34}(- \sign(\theta_{1})R (\alpha_{1}))$, where the function $\sign(x)$ equals $1$ for positive  $x$, and $-1$ otherwise.  The two partial fusions can be combined  using the rules specified in Fig.~(\ref{Identities}c).

\paragraph{Bridging} Having covered the merging procedure, we will  now describe the bridging procedure, shown in Fig.~(\ref{bridge}),  which aims to generate $U_{34}(\pm \pi /4)$.  Fig.~(\ref{bridge}a)  starts with a graph that has a cherryless tilted intercore vertex and  no pre-established graph edge between $3$ and $4$.  The rotation $M_{2}(\pm \alpha_{1})\cdot S$, will target $U_{34}(\pm \pi/4)$, where $S$ is the diagonal matrix with elements $(1,i)$.  Again the success probability is $p_{s} (\alpha_{1})$.  A failure results in a weighted edge of an angle $\mp R(\alpha_{1})$.

In parallel with the merging procedure, Figs.~(\ref{bridge}c)\&(\ref{bridge}b) show how to proceed, with another perfect intercore  vertex or another tilted intercore vertex, respectively.  In contrast  with the merging procedure, weighted edges combine by a simple rule  of addition.  Consequently, the exact amount of extra weighted edge  must be targeted.  Given a pre-existing weighted edge of angle $ \theta_{1}$ an additional $U_{34}(\pm \pi/4 - \theta_{1})$ is  required. However, when the bridging process involves a tilted  intercore vertex together with some pre-existing weighted edge (Fig.~\ref{bridge}c), calculating the required rotation is more involved.  Rotating the tilted vertex by $M(\beta_{\pm}) \cdot S $ and measuring, causes either $U_{34}(\pm \pi /4 - \theta_{1})$ or  $U_{34}(\lambda_{f})$, when:
\begin{eqnarray}
\label{BETA}
        \beta_{\pm} & = & N \cos (\alpha_{2})(\pm \cos(\theta_{1}-\sin(\theta_{1}))), \\ 
        N & = & (1 \mp \sin(2 \theta_{1} \cos (2 \alpha_{2})) )^{-\frac{1}{2}}.
\end{eqnarray}
\noindent The targeted $U_{34}(\lambda_{s})$ will be successful with  probability $p_{b}(\alpha_{2}, \theta_{1})= p_{s}(\alpha_{2})N^{2}$.  As before, this probability can be made to be always larger than $p_{s} (\alpha_{2})$, by using the sign freedom in the rotation $M(\beta_{\pm}) \cdot S$.  If the measurement fails, then a weighted edge $U_ {34}(\lambda_{f})$ is added, where:
\begin{equation}
\label{LF}
    \cos ( \lambda_{f} ) = \arccos \left( \frac{ \cos (\theta_{1}) \cos(\beta_{\pm}) } {\sqrt{1-p_{b}(\alpha_{2}, \theta_{1})}}) \right).
    \end{equation}

\paragraph{Improvements}  A rough measure of the improvement made by  our scheme can be reached by calculating the increase in success  probability for a single attempt at an entangling operation.   Consider two cavities differing in coupling strength by $10\%$ that  have the photon emission rates in Fig.~(\ref{GHZfuse}b), provided we  neglect photon loss.  In a naive scheme that post-selects projective measurements with a fidelity of less than $1-10^{-5}$, the success  probability drops from the inherent $50 \%$ to only $4\%$.  Our  scheme will attempt to bridge or merge any projective measurement  that produces a finite amount of entanglement.  A successul bridge or  merge will, in the limit of ideal detectors, generate unit fidelity  entangled states.  The overall success probability becomes $24\%$; a  substantial improvement on the naive approach by a factor of 6 \cite{CFBK01a}. 

\paragraph{Conclusions} A realistic model of distributed quantum  computing gives rise to an interesting class of random but monitored  errors, which are treated as a generalization of graph states.  A set  of strategies has been presented that adapt the growth scheme to  tackle these errors.  In some instances a failed attempt generates a  state that is described by further graph generalizations, but these  states possess partial entanglement that is recycled in later  attempts.  The benefit of the scheme is that high-fidelity graph  states can be constructed when using cavities with varying physical  parameters without suffering the severe loss in success probability  that comes with a naive post-selection strategy.

\paragraph{Acknowledgements} We would like to thank Dan Browne and  Peter Rohde for useful comments on the manuscript.  This research is  part of the QIP IRC (GR/S82176/01). SCB acknowledges  support from the Royal Society.

\end{document}